\documentclass[pdflatex,sn-vancouver-num]{sn-jnl}
\usepackage{graphicx}%
\usepackage{multirow}%
\usepackage{amsmath,amssymb,amsfonts}%
\usepackage{amsthm}%
\usepackage{mathrsfs}%
\usepackage[title]{appendix}%
\usepackage{xcolor}%
\usepackage{textcomp}%
\usepackage{manyfoot}%
\usepackage{booktabs}%
\usepackage{algorithm}%
\usepackage{algorithmicx}%
\usepackage{algpseudocode}%
\usepackage{listings}%
\usepackage{makecell}
\usepackage{arydshln}
\usepackage{hyperref}

\begin{document}

\title[Temporal Trends in Incidence of Dementia in a Birth Cohorts Analysis of the Framingham Heart Study]{Temporal Trends in Incidence of Dementia in a Birth Cohorts Analysis of the Framingham Heart Study}

\author[1,2]{\fnm{Paula} \sur{Staudt}}
\equalcont{These authors contributed equally to this work.}

\author[1]{\fnm{Anika} \sur{Schlosser}}
\equalcont{These authors contributed equally to this work.}

\author[3]{\fnm{Annika} \sur{Möhl}}

\author[1,2]{\fnm{Martin} \sur{Schumacher}}

\author*[2,4]{\fnm{Nadine} \sur{Binder}}\email{nadine.binder@uniklinik-freiburg.de}

\affil[1]{\orgdiv{Institute of Medical Biometry and Statistics}, \orgname{Faculty of Medicine and Medical Center, University of Freiburg}, \orgaddress{\city{Freiburg, Germany}}}

\affil[2]{\orgdiv{Freiburg Center for Data Analysis, Modeling and AI}, \orgname{University of Freiburg}, \orgaddress{\city{Freiburg, Germany}}}

\affil[3]{\orgdiv{Institute of Medical Biometry and Epidemiology}, \orgname{University Medical Center Hamburg-Eppendorf}, \orgaddress{\city{Hamburg, Germany}}}

\affil[4]{\orgdiv{Institute of General Practice/Family Medicine}, \orgname{Medical Center and Faculty of Medicine, University of Freiburg},\orgaddress{ \city{Freiburg, Germany}}}



\abstract{
\textbf{Background:}
Dementia leads to a high burden of disability and the number of dementia patients worldwide doubled between 1990 and 2016. Nevertheless, some studies indicated a decrease in dementia risk which may be due to a bias caused by conventional analysis methods that do not adequately account for missing disease information due to death.

\textbf{Methods:}
This study re-examines potential trends in dementia incidence over four decades in the Framingham Heart Study. We apply a multistate modeling framework tailored to interval-censored illness–death data and define three non-overlapping birth cohorts (1915-1924, 1925-1934, and 1935-1944). Trends are evaluated based on both dementia prevalence and dementia risk, using age as the underlying timescale. Additionally, age-conditional dementia probabilities stratified by sex are estimated.

\textbf{Results:} 
A total of 731 out of 3828 individuals were diagnosed with dementia. The multistate model analysis revealed no temporal decline in dementia risk across birth cohorts, irrespective of sex. When stratified by sex and adjusted for education, women consistently exhibited higher lifetime age-conditional risks (46\%-50\%) than men (30\%-34\%) over the study period.

\textbf{Conclusions:}
We recommend using a combination of multistate approach and separation into birth cohorts to adequately estimate trends of disease risk in cohort studies as well as to communicate patient-relevant outcomes such age-conditional disease risks.}

\maketitle

\section*{Introduction}\label{Indroduction}
Dementia leads to a high degree of disability among those affected, which is reflected in the number of DALYs attributable to the disease \cite{nichols_global_2019}. The number of dementia patients worldwide doubled between 1990 and 2016, which can mainly be explained by the aging population \cite{nichols_global_2019}. In contrast to this trend, there are indications for a possible decrease in the dementia incidence, especially in high-income countries \cite{prince_recent_2016, roehr_is_2018}. The current relevance of the topic is particularly highlighted by the Dementia Risk Prediction Project (DRPP) \cite{krefman_cohort_2024}, which harmonizes data from 16 international cohorts comprising over 119000 participants and make them available for analysis. Crucial for the analysis of such datasets, however, is which statistical models are used.

Most studies on dementia incidence use conventional methods such as Cox models \cite{roehr_is_2018, whiteley_long-term_2021, wolters_twenty-seven-year_2020, chibnik_trends_2017} focusing on dementia only, as recently again demonstrated by Yan et al. \cite{yan_coffee_2025} and Ding et al. \cite{ding_secular_2025}. However, the multistate approach is a more suitable method to derive valid estimates \cite{binder_multi-state_2019, chen_dementia_2023}, which has recently also been acknowledged by other work in dementia research \cite{le-rademacher_utility_2022, weisbach_left-censored_2021, doblhammer_declining_2025}. A prominent example of a study that found a linear decline of dementia incidence over time is the study of Satizabal et al. \cite{satizabal_incidence_2016}, who analyzed Framingham Heart Study (FHS) data constructing four non-overlapping 5-year epochs ranging from 1977 to 2008. Two recently published studies, that also used the division into epochs, came to similar results of declining dementia incidences \cite{farina_trends_2022, johnsen_incidence_2023}. Declining incidences were also described in a systematic review of cohort studies published in 2024, which also included Satizabal's study \cite{mukadam_changes_2024}.
However, the interpretation of temporal trends critically depends on the accuracy of both disease and mortality ascertainment. A notable example of an intensive mortality ascertainment effort is the study by Schramm et al. \cite{schramm_incidence_2025}, which utilized death certificates together with supplementary medical documentation. Despite these efforts, data quality remained constrained by limited document availability, and without the application of a multistate model, potential disease onset between the last assessment and death could not be accounted for.
This unobserved interval, known as missing disease information due to death (MDID) \cite{binder_missing_2014}, may contribute to an apparent decline in dementia incidence. As Nichols et al. \cite{nichols_estimating_2025} have shown, selective mortality between observation waves can lead to an underestimation of disease trajectories, since individuals who die between measurement points are not random. These findings underscore the importance of explicitly accounting for MDID in studies of dementia incidence. We have recently addressed this issue using a multistate model-based approach that allows adjustment for the probability of disease onset after the last observation and did not find convincing evidence of a decline in dementia incidence over the four epochs \cite{binder_multi-state_2019}.
Another source of bias in analyzing dementia trends could be the actual division of cohorts into epochs, as it was also done by Satizabal et al. Age and cohort effects can be better separated by dividing the study population into birth cohorts rather than epochs \cite{grodstein_trends_2023, bruck_projected_2022}. The additional use of age as time scale in the analyses allows a natural interpretation of the results, i.e. the risk of developing a disease at a certain age \cite{le-rademacher_utility_2022}.

In this work, using the FHS data, we combined the above-mentioned design and analysis aspects by creating non-overlapping birth cohorts and applying the multistate approach in order to re-examine the trend in dementia incidence in the FHS cohort. Furthermore, we provided patient-relevant outcomes (the risk of developing dementia at a certain age) and were able to evaluate temporal trends. 

\section*{Methods}\label{Methods}
\subsubsection*{The Framingham Heart Study}\label{The_Framingham}
The FHS is a community-based, longitudinal cohort study initiated in 1948 in Framingham, Massachusetts, primarily to investigate risk factors for cardiovascular disease. The study initially recruited 5209 residents who were 28 to 74 years old in 1948 (original cohort). Participants were followed up every 2 years for detailed examinations, resulting in a total of 32 visits ending in 2014, with event follow-up through 2018. In 1971, 5124 offspring of the original cohort and their spouses, aged 5 to 70 years, were sampled for the offspring cohort. Participants in this cohort have completed up to 10 examinations every 4 years until 2021. 
Monitoring of cognitive status began in 1975 for the original cohort and in 1979 for the offspring cohort. Specifically, participants have been assessed using the Mini-Mental State Examination (MMSE) at the examinations since 1981 and 1991 respectively. Participants identified as having possible cognitive impairment based on performance on these screening assessments or on references from family and physicians are invited to undergo additional neurological and neuropsychological testing. Each case of possible cognitive decline was reviewed by a dementia review panel to determine the date of dementia onset using data from previously performed serial neurological and neuropsychological assessments, telephone interviews with caregivers, medical records and neuroimaging studies. Early reviews prior to 2001 were repeated to meet current diagnostic criteria. After the death of a participant, review efforts were undertaken to assess a possible cognitive decline based on medical and nursing records. Detailed information about the dementia screening process in the FHS has been published previously \cite{satizabal_incidence_2016, seshadri_lifetime_1997, seshadri_operationalizing_2011}. Data on dementia review and survival was provided until 2018 for the original cohort. For the offspring cohort survival data were available through 2021, where dementia data where available through 2022. Educational attainment information was provided through a validated dataset covering data up to 2011. Further details on the dataset can be found in the \textit{Supplementary Material}.

\subsubsection*{Creation of analysis datasets}\label{Creation_of}
The analyses were based on the data extract provided to us in 2025. We received data from a total of 5209 subjects from the original cohort, and data from 5124 subjects from the offspring cohort. To disentangle age and cohort effects, we constructed three sequential, non-overlapping birth cohorts of 10 years each, consisting of participants from the original cohort and the offspring cohort born in the years 1915-1924 (birth cohort 1), 1925-1934 (birth cohort 2) and 1935-1944 (birth cohort 3), respectively. For each birth cohort, we included participants with at least one dementia-free screening at or after the age of 60 (known to be free of dementia at the age of 60). As monitoring of cognitive status began in 1975, participants born before 1915 could not have attended a dementia assessment before the age of 60 and were therefore not included in the analysis sample. In addition, participants born after 1945 were not included because they were less than 77 years old at the end of follow-up in 2021 and the aim is to use age as the time scale for the analysis starting at age 60. We constructed an indicator for the onset of dementia based on the reviews and the results of the neurological tests. Detailed information on the construction is provided in Figure S1 and Table S1 of the \textit{Supplementary Material}. Education is considered as a dichotomized variable that distinguishes between low and high educational attainment. The former describes having attained at most a high school diploma, i.e. having completed up to 12 years of schooling. The latter includes all tertiary education, i.e. at least attending college or obtaining an academic degree. Detailed information on the variable describing educational attainment can be found in Figure S2 of the \textit{Supplementary Material}.

\begin{figure}[t]
\centering
\includegraphics[width=0.6\textwidth]{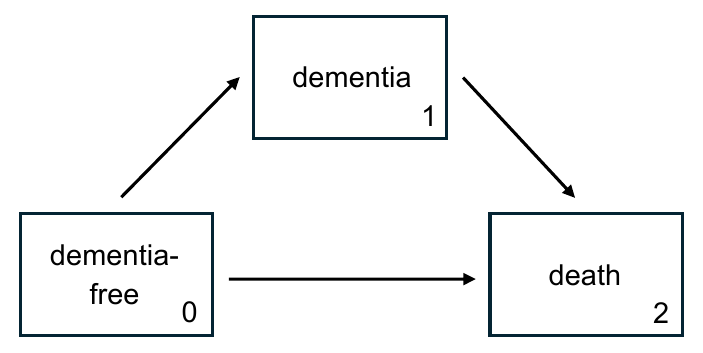}
\caption{Illness-death model}
\label{fig1}
\end{figure}

\subsubsection*{Statistical analysis}\label{Statistical_analysis}
Using a multistate model approach, we calculated dementia prevalence and dementia risk for each birth cohort adjusted for sex. Age was used as the time scale. Dementia prevalence is defined as the probability of having dementia at a given age conditional on being alive and free of dementia at age 60. Dementia risk is the cumulative probability of developing dementia at any time between age 60 and a specified age, conditional on being alive and free of dementia at age 60. Estimates are derived from the spline-based penalized likelihood implemented in the R package SmoothHazard \cite{touraine_smoothhazard_2017} which is fitted based on the incomplete interval-censored multistate data. Details on model estimation and smoothing specification are available in the \textit{Supplementary Material}. The underlying multistate model, represented in Figure \ref{fig1}, describes the transitions of a subject or a group through different states. The so-called illness-death model features three states: state 0 “dementia-free”; state 1 “dementia”, and state 2 “death”. Each individual starts in state 0 and may remain in this state until the end of the observation period or until they transition to state 1 or state 2 (either directly or via state 1), as indicated by the arrows in the model. The transition to state 1 may be interval-censored for some subjects, meaning the time of dementia onset is not known exactly but occurred between two time points, e.g., examinations. Taking into account interval-censored data is important to eliminate the problem of potential missing disease information due to death and due to loss-to-follow-up which are critical in cohort studies with long follow-up intervals and diseases with differential mortality \cite{joly_penalized_2002}. Participants were censored at the last time they were known to be alive and dementia-free, that is the last contact with the study. Since the role of educational status in the development of dementia is not fully understood \cite{robitaille_transitions_2018}, education was considered as covariate to account for the association of educational level and the risk of developing dementia. Age-conditional probabilities are included in the analysis as a patient-oriented quantity to predict the risk of dementia in the next 10 or 20 years, or ever in life, conditional on being free of dementia at a given age.

\begin{figure}[t]
\centering
\includegraphics[width=0.86\textwidth]{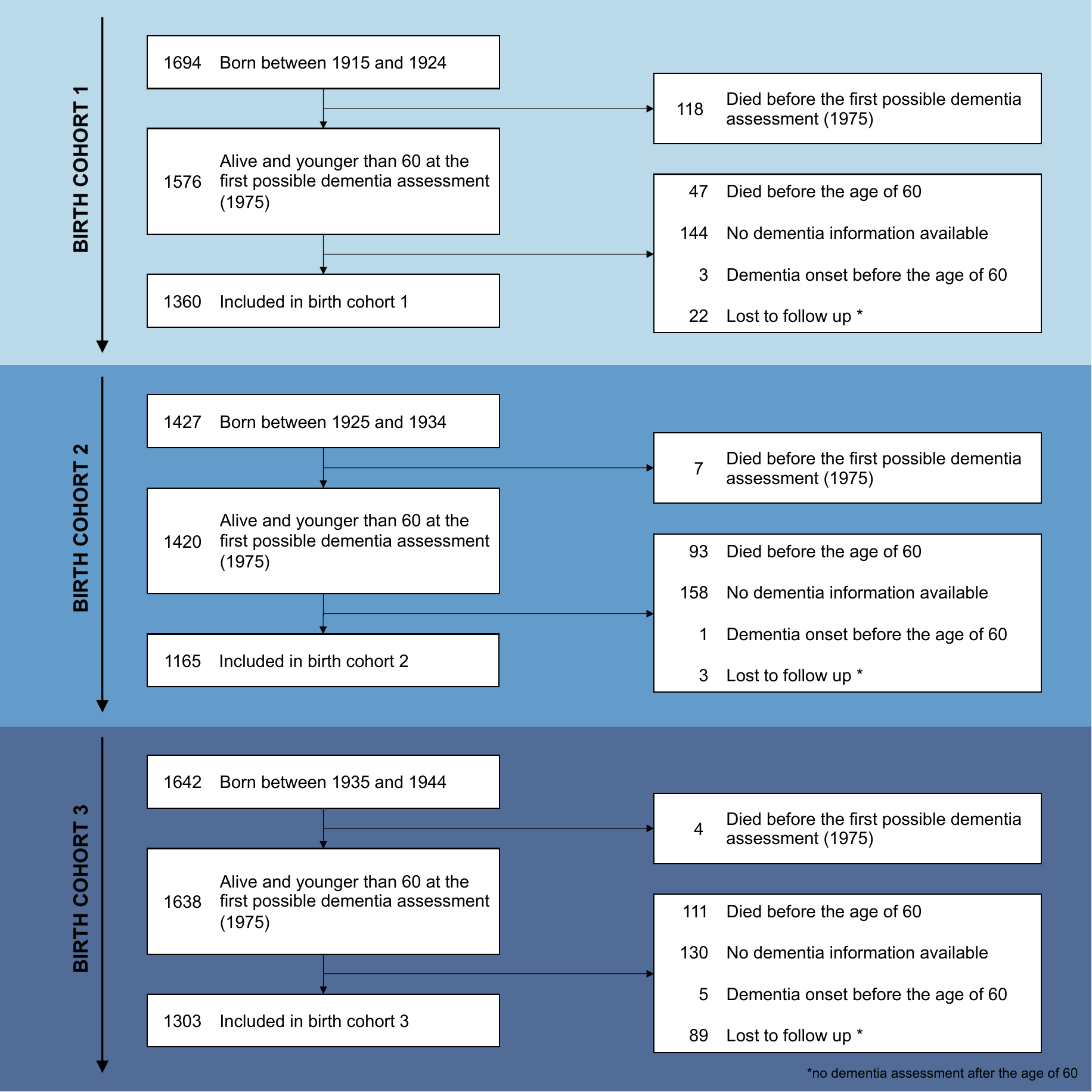}
\caption{Construction of birth cohorts}
\label{fig2}
\end{figure}

\section*{Results}\label{Results}
The final analysis sample comprises 3828 individuals selected from the initial 10333 individuals in the original cohort and the offspring cohort as follows. First, 251 individuals were excluded due to absence of matching Exam 1 record, preventing calculation of their year of birth. Next, 4048 participants were born before 1915 and 1271 after 1944 and therefore have to be excluded from the sample. The remaining sample across the three birth cohorts consists of 4763 participants. In addition, all participants who died before 1975, i.e., before the first dementia assessment, have to be excluded from the sample. Naturally, this applies to a significantly higher number of participants born between 1915 and 1924 compared to the other two birth cohorts. In order to include only participants who are free of dementia at the age of 60, participants must be excluded if any of the following four reasons applies to them: They died before the age of 60, information on dementia is not available, dementia was diagnosed before the age of 60, or they did not undergo a dementia assessment after the age of 60. A flowchart describing the numbers per exclusion criteria for each birth cohort is depicted in Figure \ref{fig2}. 
The resulting three birth cohorts are similar in size, consisting of 1360 (birth cohort 1), 1165 (birth cohort 2) and 1303 (birth cohort 3) individuals respectively. In each cohort, the percentage of female subjects is slightly predominant at about 53\% (see Table S2). Educational attainment was documented for 3676 out of the 3828 subjects (96\%) in our analysis sample. The educational level rises over the birth cohorts, as the percentage of cohort members with higher education increases from 34.8\% in the first birth cohort to 65.2\% in the third birth cohort (see Table S2). 

Table \ref{table1} shows the number of individuals in each birth cohort for each possible transition in the illness-death model as depicted in Figure \ref{fig1}, with an additional focus on missing disease information due to death (i.e., death and dementia inconclusive) and due to loss-to-follow up (i.e., alive and dementia inconclusive). 

\begin{table}[t]
\centering
\small
\caption{Numbers and proportions in percent for possible states or transitions in the illness-death model per birth cohort with an additional focus on missing disease information due to death (i.e., death and dementia inconclusive) and due to loss-to-follow up defined as `no dementia assessment for 4 years' (i.e., alive and dementia inconclusive)}
\label{table1}
\begin{tabular}{lrcrcrcrc}
\hline
& \multicolumn{2}{c}{\textbf{Total}} & \multicolumn{2}{c}{\textbf{Birth cohort 1}} & \multicolumn{2}{c}{\textbf{Birth cohort 2}} & \multicolumn{2}{c}{\textbf{Birth cohort 3}} \\
& Count & \% & Count & \% & Count & \% & Count & \% \\
\hline
\textbf{No. of subjects} & 3828 &  & 1360 &  & 1165 &  & 1303 &  \\
\hline
\makecell[l]{\textbf{Alive and}\\\textbf{dementia-}\\\textbf{free}} & 719 & 18.8\% & 14 & 1.0\% & 160 & 13.7\% & 545 & 41.8\% \\
\specialrule{0.3pt}{0pt}{0pt}
\makecell[l]{\textbf{Alive and}\\\textbf{dementia}\\\textbf{inconclusive}} & 578 & 15.1\% & 22 & 1.6\% & 191 & 16.4\% & 365 & 28.0\% \\
\specialrule{0.3pt}{0pt}{0pt}
\makecell[l]{\textbf{Diagnosed}\\\textbf{dementia}} & 731 & 19.1\% & 427 & 31.4\% & 228 & 19.6\% & 76 & 5.9\% \\
\specialrule{0.3pt}{0pt}{0pt}
\makecell[l]{\textbf{Death}\\\textbf{without}\\\textbf{dementia}} & 1276 & 33.3\% & 667 & 49.1\% & 399 & 34.2\% & 210 & 16.1\% \\
\specialrule{0.3pt}{0pt}{0pt}
\makecell[l]{\textbf{Death after}\\\textbf{dementia}\\\textbf{diagnosis*}} & 662 & 90.6\% & 423 & 99.1\% & 194 & 85.1\% & 45 & 59.2\% \\
\specialrule{0.3pt}{0pt}{0pt}
\makecell[l]{\textbf{Death and}\\\textbf{dementia}\\\textbf{inconclusive}} & 524 & 13.7\% & 230 & 16.9\% & 187 & 16.1\% & 107 & 8.2\% \\
\hline
\end{tabular}
\footnotesize
\textit{*Death after dementia diagnosis is a subgroup of diagnosed dementia cases.}
\end{table}

\begin{figure}[t]
\centering
\includegraphics[width=0.97\textwidth]{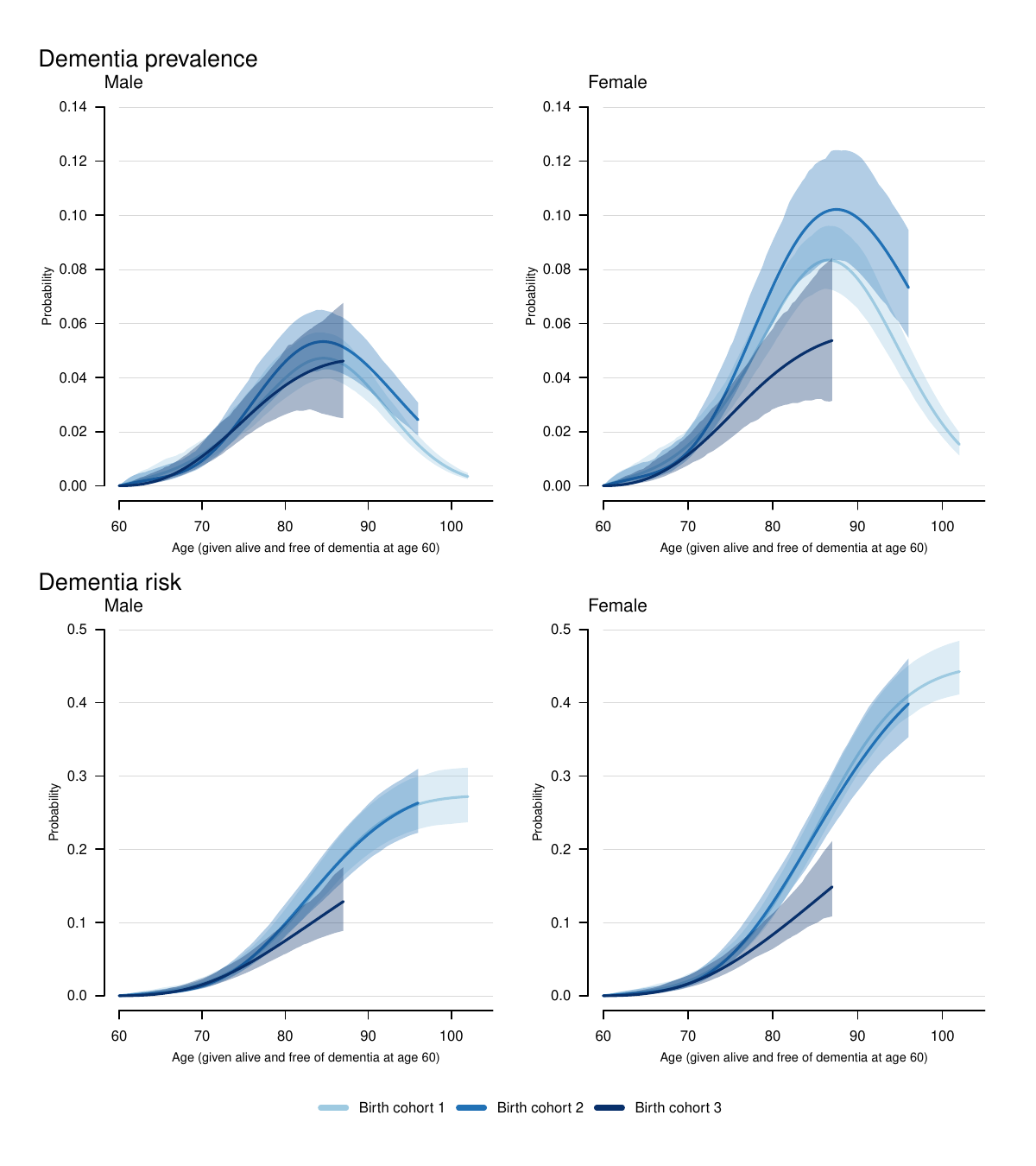}
\caption{Estimated dementia prevalence (upper panel) and estimated dementia risk (lower panel) stratified by birth cohort and adjusted for sex together with their 95\% confidence intervals}
\label{fig3}
\end{figure}

A total of 731 confirmed dementia cases were observed across the three birth cohorts, with a natural decline over the birth cohorts. The drastic increase in the number of participants who are alive but who have not attended an assessment in the last four years is noteworthy emphasizing the importance of accounting for interval-censored data. A graphical representation of Table \ref{table1} showing the status information per birth year can be found in Figure S3 of the \textit{Supplementary Material}. Extended versions of this table can also be found in the \textit{Supplementary Material} showing the numbers for each possible transition in the illness-death model additionally stratified by sex(Table S3) and education(Table S4). Figure \ref{fig3} illustrates the estimated dementia prevalence (upper panel) and estimated dementia risk (lower panel) stratified by birth cohort and adjusted for sex together with their 95\% confidence intervals (CI).

Dementia prevalence is significantly higher for women than for men. In the first and the second birth cohort dementia prevalence reaches its maximum around age 85 in men and 87-88 in women. The maximum in the second birth cohort is notably higher than in the first birth cohort for men and for women. The third birth cohort shows comparable dementia prevalence in men, but lower prevalence in women compared to both earlier cohorts. However, due to the wide and overlapping confidence intervals, especially at older ages, where data become sparse, it cannot be determined whether this observed difference is statistically significant. The estimated risk of dementia is substantially higher for women than for men. When comparing the first two birth cohorts, the risk curves are nearly identical for both sexes, showing no evidence of decline. The last birth cohort shows a reduction in risk compared to both earlier cohorts. However, because of the small number of events and the short observation period in this cohort, the precision of our estimates is limited, longer follow-up will be needed before drawing firm conclusion. Considering education as a covariate in the multistate model analysis, higher educational attainment was associated with a lower risk of developing dementia (hazard ratio (HR) and 95\% CI = 0.85 [0.73, 0.98]) and a substantially reduced risk of direct death (HR and 95\% CI = 0.70 [0.63, 0.78]). However, higher education did not significantly influence mortality after dementia onset (HR and 95\% CI = 1.00 [0.96, 1.04]). 

With no observed significant difference in dementia risk between birth cohorts, the analysis sample was further used to estimate age-conditional dementia probabilities stratified by sex and adjusted for education. The results of this patient-oriented quantity for predicting the risk of dementia in the next 10 or 20 years, or ever in life, conditional on being free of dementia at a given age are presented in Table \ref{table2}. Age-conditional probabilities increase sharply with advancing age for both sexes. However, women have significantly higher 20-year and lifetime risks (approximately 46\%-50\% risk at ages 60-80) than men (approximately 30\%-34\% risk). In contrast, 10-year risks are similar for women and men at younger baseline ages.

\begin{table}[t]
\centering
\caption{Age-conditional dementia probabilities stratified by sex and adjusted for education}
\label{table2}
\begin{tabular}{lccc}
\hline
Men at the age of \quad  \quad & 10 years & 20 years & Ever \\
\hline
60 & 1.93\% & 9.4\% & 30.15\% \\
65 & 4.23\% & 16.69\% & 31.25\% \\
70 & 8.62\% & 24.66\% & 32.48\% \\
75 & 15.57\% & 30.75\% & 33.66\% \\
80 & 23.21\% & & 34.41\% \\
\hline
\end{tabular}
\begin{tabular}{lccc}
\hline
Women at the age of & 10 years & 20 years & Ever \\
\hline
60 & 1.08\% & 10.3\% & 46.2\% \\
65 & 3.91\% & 20.54\% & 46.9\% \\
70 & 9.87\% & 32.79\% & 48.26\% \\
75 & 19.22\% & 43.04\% & 49.57\% \\
80 & 29.97\% & & 50.08\% \\
\hline
\end{tabular}
\end{table}

\section*{Discussion}\label{Dis}
The present work was designed to re-examine a potential trend in dementia incidence over the past 40 years in the FHS using an alternative design and adequate methodology. For this purpose, we constructed three non-overlapping birth cohorts and chose a multistate model, designed for interval-censored illness-death type data. Specifically, we considered the natural time scale "age" and calculated age-conditional risks of dementia that are informative measures in terms of epidemiology. A total of 731 dementia cases were observed across the three birth cohorts. The curves in Figure \ref{fig3} indicate a slight, but non-significant trend for a decline of dementia risk in the youngest birth cohort. Overall, our results do not support the conclusion that dementia incidence in the Framingham cohort declined over the specified time period. The age-conditional probabilities for the development of dementia seem to be high in our work. However, Ben-Hassen et al. showed similarly high age-dependent dementia risks in their work \cite{ben-hassen_five-year_2021}. Importantly, this study focused on the methodological approach rather than on identifying risk factors. Although we examined differences by sex and education, both well-established risk factors, the proposed modeling approaches naturally allow for the inclusion of additional variables. 

\subsubsection*{Strengths and limitations} \label{Stren}
The FHS is characterized by its long follow-up time and longitudinal data from a large cohort in several generations. The screening for dementia and monitoring of cognitive status since 1975 allowed the analysis of dementia incidence over time \cite{satizabal_incidence_2016, farmer_neuropsychological_nodate}.  The strengths of our study can be summarized by the following major aspects concerning the design, methods and measurement of the outcome. First, on the design level, we constructed birth cohorts instead of separating data by epochs. Non-overlapping birth cohorts are clearly separated, meaning that a subject can only appear in one cohort. When separating by epochs, one and the same subject can appear in more than one epoch (e.g. a person being dementia free in epoch 1 will also appear in epoch 2), diluting the effect. Creating birth cohorts enables to properly disentangle age effects and potential temporal trends \cite{tom_association_2020}. Second, on the methods level, we applied the multistate model approach, which is well suited for analyzing time-to-event data, as they offer a more flexible framework than conservative methods \cite{le-rademacher_utility_2022}. Moreover, it allows to adequately account for potential missing disease information due to death. Third, on the measurement level, using age as the timescale is intuitive and helps to provide estimates in a patient-relevant way. The age-conditional risk of dementia is easily interpretable and can be compared with results from similar studies. The three above-mentioned methodological issues are already (partially) used in the literature, but often not in combination. For example, although the creation of birth cohorts is used by Tom et al., conservative Cox proportional hazard models were used \cite{tom_association_2020}. Therefore, it is likely that a potentially observed decline in dementia incidence is not primarily due to improvements in education or lifestyle but due to the lack of the methodological aspects mentioned in our work \cite{farina_trends_2022,  johnsen_incidence_2023, mukadam_changes_2024}.

The FHS only includes subjects living in the U.S. but of European ancestry, limiting the generalizability of the study results.
Regarding the methods applied, the penalized likelihood approach makes assumptions, e.g. on the shape of splines, to provide valid estimates. Even if there is a potential risk for incorrectness of these assumptions, the approach is more suitable to estimate the risk of dementia than naive modeling approaches censoring death. Alternatively, the multiple imputation approach, which we have not discussed in detail in this paper, can be used.

\subsubsection*{Conclusion} \label{Con}
Besides highlighting challenges in analyzing incidence trends, we proposed a more suitable approach than conventional methods censoring death in order to estimate the risk of dementia. Since the problem of bias due to missing disease information due to death is also relevant for other cohorts, we hope that the combination of multistate approach and separation into birth cohorts will be applied more frequently in future analyses.

\bibliography{bib_neu}


\subsection*{Funding}
The work of NB has been funded in part by the Deutsche Forschungsgemeinschaft (DFG, German Research Foundation) – Project-ID 499552394 – SFB 1597 (https://www.dfg.de/en). 
The funding agreement ensured the authors’ independence in designing the study, collecting, analyzing, and interpreting the data, writing of the manuscript, and in the decision to submit the manuscript for publication.

\subsection*{Author Contributions}
The concept and overall aims of the manuscript were developed through discussions between MS and NB. PS and AS prepared the analysis dataset and conducted all statistical analyses, including the presentation of results, under the supervision of NB. AS and NB drafted the manuscript, with AM contributing to the introduction and discussion sections. PS refined and streamlined the full draft. All authors critically reviewed successive versions of the manuscript and approved the final version.

\section*{Ethics Declarations}
\subsection*{Competing Interests}
The authors declare no competing interests.

\end{document}